\documentclass{aa}

\usepackage{graphicx}
\usepackage[varg]{txfonts}
\usepackage[colorlinks=true,linkcolor=blue,citecolor=blue, urlcolor=blue]{hyperref}
\usepackage{appendix}

\usepackage{graphicx}	
\usepackage{amsmath}	
\usepackage{subfigure}
\usepackage{caption}
\usepackage{subcaption}
\usepackage{multirow}
\usepackage{pifont}
\usepackage[normalem]{ulem}
\usepackage{booktabs}
\usepackage{sidecap}

\defcitealias{Nepal2024}{N24}

\begin{document}

    \title{The oldest Milky Way stars: New constraints on the age of the Universe and the Hubble constant}

   \author{Elena Tomasetti\inst{1,2} \fnmsep\thanks{\email{elena.tomasetti2@unibo.it}}
          \and
           Cristina Chiappini\inst{3}
           \and
           Samir Nepal\inst{3,4}
          \and
           Michele Moresco\inst{1,2}
           \and
           Carmela Lardo\inst{1,2}
           \and 
           \\Andrea Cimatti\inst{1,2}
           \and
           Friedrich Anders\inst{5,6,7}
           \and
           Anna B. A. Queiroz\inst{8,9}
           \and
           Guilherme~Limberg\inst{10}
           }

   \institute{
            Dipartimento di Fisica e Astronomia “Augusto Righi”–Università di Bologna, via Piero Gobetti 93/2, I-40129 Bologna, Italy
            \and
            INAF - Osservatorio di Astrofisica e Scienza dello Spazio di Bologna, via Piero Gobetti 93/3, I-40129 Bologna, Italy
            \and
            Leibniz-Institut für Astrophysik Potsdam (AIP), An der Sternwarte 16, 14482, Potsdam, Germany
            \and
             Institut für Physik und Astronomie, Universität Potsdam, Karl-Liebknecht-Str. 24/25, 14476, Potsdam, Germany
            \and
            {Departament de Física Quàntica i Astrofísica (FQA), Universitat de Barcelona (UB), Martí i Franquès 1, 08028 Barcelona, Spain}
            \and{Institut de Ciències del Cosmos (ICCUB), Universitat de Barcelona (UB), Martí i Franquès 1, 08028 Barcelona, Spain}
            \and{Institut d'Estudis Espacials de Catalunya (IEEC), Edifici RDIT, Campus UPC, 08860 Castelldefels (Barcelona), Spain}
            \and
            Instituto de Astrof\'isica de Canarias, E-38200 La Laguna, Tenerife, Spain
            \and
            Departamento de Astrof\'isica, Universidad de La Laguna, E-38205 La Laguna, Tenerife, Spain
            \and
            Kavli Institute for Cosmological Physics, University of Chicago, 5640 S. Ellis Avenue, Chicago, IL 60637, USA
             }

    \titlerunning{The oldest Milky Way stars}
    \authorrunning{E. Tomasetti et al.}
 
  \abstract
   {}
   {In this work, we exploit the most robust, old, and cosmology--independent age estimates of individual stars from \textit{Gaia} DR3 to place a lower bound on the age of the Universe, $t_U$. These constraints can be used as an anchor point for any cosmological model, providing an upper limit to the Hubble constant $H_0$.}
   {We consider the stellar age catalog of \citet{Nepal2024}, selecting 3,000 of the oldest and most robustly measured main sequence turn-off (MSTO) and subgiant branch (SGB) stars, with ages older than 12.5 Gyr and associated uncertainty below 1 Gyr. Stellar ages are derived via isochrone fitting using the Bayesian code \texttt{StarHorse}, spanning the uniform range 0 -- 20 Gyr, not considering any cosmological prior knowledge on $t_U$. Applying a conservative cut in the Kiel diagram, strict quality cuts on both stellar parameters and posterior probability distribution shapes, and filtering out potential contaminants, we isolate a final sample of 160 bona-fide stars, representing the most numerous sample of precise and reliable MSTO and SGB stars ages available to date.}
   {The age distribution of the final sample peaks at $13.6 \pm 1.0\,\text{(stat)} \pm 1.3\,\text{(syst)}$ Gyr. Assuming a maximum formation redshift for these stars of $z_f = 20$, corresponding to a formation delay of $\sim$0.2 Gyr, we obtain a lower bound on $t_U$ of $\mathrm{t_U \geq 13.8 \pm 1.0\,\text{(stat)} \pm 1.3\,\text{(syst)}}$ Gyr. Considering the $10^{th}$ percentile of the posterior probability distributions of the individual stars, we find that, at 90\% CL, 70 stars favour $t_U > 13$ Gyr, while none exceeds 14.1 Gyr. An oldest age younger than 13 Gyr for this sample is incompatible with the data, even considering the full systematic error budget.}
   {This work presents the first statistically significant use of individual stellar ages as cosmic clocks, opening a new, independent approach for cosmological studies. While this analysis represents already a significant step forward, future \textit{Gaia} data releases will enable even larger and more precise stellar samples, further strengthening these constraints.}


   \maketitle
%

\section{Introduction}
\label{sec:1Intro}

\begin{figure*}
    \centering
    \includegraphics[width=\linewidth]{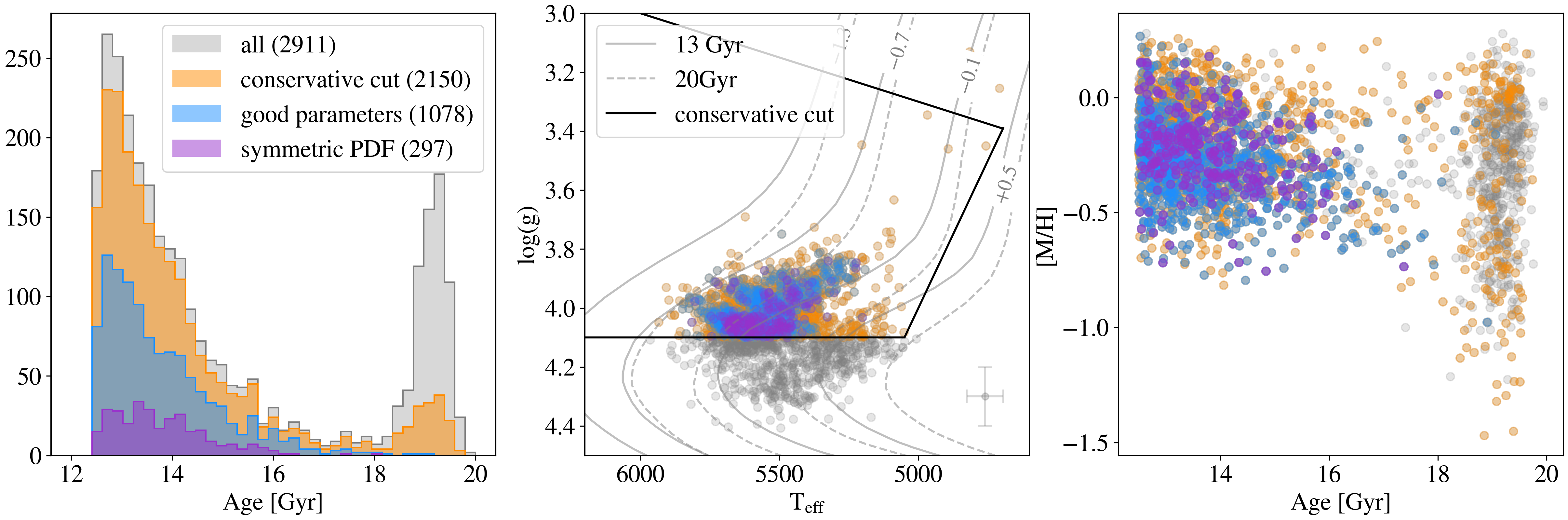}
    \caption{Age distribution (left), Kiel diagram (center), and age-metallicity coverage (right) for each step of the selection process, before visual inspection. In the central panel, PARSEC isochrones at different ages and metallicities are shown in grey, and at the bottom left, the average error in $\mathit{log(g)}$ and $\mathit{T_{eff}}$ for the full sample is shown. The peak at 19 Gyr visible in the histogram is due to contaminants, as explained in Sect. \ref{sec:selectionprocess} and Appendix \ref{appendix_conscut}.}
    \label{fig:3histo_kiel}
\end{figure*}

Although the $\Lambda$CDM model is widely successful in modern cosmology in describing several observables, recent high-precision measurements from key probes have highlighted discrepancies between early- and late-Universe constraints. One of the critical parameters is the local expansion rate of the Universe (the Hubble constant, $H_0$) that shows a tension of 4-5$\sigma$ when measured from different probes \citep[Cepheids and SNe as opposed to CMB, see, e.g.,][]{Abdalla2022,Kamionkowski2023}; this could hint to new physics or hidden systematics in the data, emphasising the need for complementary methods to trace the Universe's expansion history and uncover the source of this tension.

The absolute ages of the oldest objects at $z=0$ are critical in cosmology, as they can set a lower bound on the current age of the Universe \citep[$t_U$,][]{Krauss2003}. Assuming a cosmological model, this lower limit on $t_U$ can be translated into an upper limit on the Hubble constant ($H_0$). In particular, considering the two primary independent measurements that originated the "Hubble tension", $H_0$ = 67.4 $\pm$ 0.5 km/s/Mpc \citep{PlanckCollaboration2020} and $H_0$ = 73.04 $\pm$ 1.04 km/s/Mpc \citep{Riess2022}, these correspond (in a flat $\Lambda$CDM model with $\Omega_m$ = 0.3) to $\mathit{t_U}$ $= 14.0 \pm 0.1$ Gyr and $t_U$ $= 12.9 \pm 0.2$ Gyr, respectively. Therefore, measuring $t_U$ with an accuracy of about 10\% can provide independent and crucial insights into this subject.

In recent years, interest has been growing in the use of stellar ages as promising cosmological probes \citep{OMalley2017, Jimenez2019, Valcin2020, Valcin2021, DiValentino2021a,BoylanKolchin2021,Moresco2022,Vagnozzi2022,Tomasetti2025,Valcin2025}, independent of the standard ones and of any cosmological model. Various methods and types of objects have been employed for this purpose, most notably isochrone fitting, applied to globular clusters (GCs) and individual stars \citep[e.g.][]{Valcin2025}, as well as full-spectrum fitting for GCs \citep{Tomasetti2025} and techniques based on white dwarf cooling sequences and nucleochronometry \citep[see][for a collection of different approaches]{Cimatti2023}. Thanks to the tremendous increase in the quality and statistics of the data for field stars in the Gaia era, very high precision is currently achievable in measuring age. Accuracy, on the other hand, represents a persistent challenge due to the presence of systematic uncertainties \citep{Soderblom2010,Valcin2021,Joyce2023}, mainly arising from stellar models' dependencies, typically dominating over the internal precision of each method. Moreover, the previously mentioned studies generally relied on small samples (a few tens of old GCs or a handful of single stars) as precise age estimates were available for a few objects, limiting the statistical robustness of the resulting cosmological inferences.

Nowadays, precise age determinations can be obtained for main-sequence turnoff (MSTO) and subgiant branch (SGB) stars, via isochrone fitting, thanks to the very high-quality data and stellar parameters obtained with cutting-edge facilities like the ESA Gaia \citep{Collaboration2016}. In this paper, we take advantage of the high-quality age measurements obtained in \citet[][N24 hereafter]{Nepal2024} for a sample of about 200’000 MSTO and SGB stars, with extremely precise parallaxes ($<$1\%) and extinction uncertainties ($<$0.2 mag). Extending the methodology adopted in \citetalias{Nepal2024}, where stellar ages were constrained by a cosmological prior of 13.73 Gyr, the ages adopted here were derived without any such upper bound, spanning the full range of the isochrone models (0.025--20 Gyr). This unprecedented dataset enables us to use stellar age dating as a cosmological probe, for the first time, with both high statistical power and internal homogeneity, while also enabling a detailed assessment of the systematic uncertainties affecting age measurements. 

In Sect. \ref{sec:2DATA} we present the dataset and the method used to derive ages, in Sect. \ref{sec:Analysis} the selection of the optimal age sample, together with the analysis of the systematic effects. In Sect. \ref{sec:4CosmoAnalysis} we report the cosmological analysis and in Sect. \ref{sec:5CONCLUSIONS} we draw our conclusions. In this work, unless stated otherwise, we assumed a flat $\Lambda$CDM cosmology with $\Omega_m$=0.3.


\section{Data}
\label{sec:2DATA}

Our study is based on a sample of 202,384 stars presented in \citetalias{Nepal2024}, soon become publicly available (Nepal et al., \textit{in prep.}), identified as the ``age sample'', where ages are derived using the Bayesian isochrone fitting code \texttt{StarHorse} \citep{Santiago2016,Queiroz2018, Queiroz2023}. The \texttt{StarHorse} code estimates stellar ages, extinctions, and distances by comparing observed data to the stellar evolutionary models from the PAdova and TRiestre Stellar Evolution Code \citep[PARSEC,][]{Bressan2012}. In particular, the main observables used as input are the stars' atmospheric parameters, photometric magnitudes, and parallaxes. Specifically, atmospheric parameters such as effective temperature ($\mathrm{T_{eff}}$), surface gravity (log(g)), and overall metallicity ([M/H]) are derived in \citet{Guiglion2024} analysing spectra from the Radial Velocity Spectrometer (RVS) using a hybrid convolution neural network (\texttt{hybrid--CNN}). The photometric magnitudes G, Bp, and Rp are from the third data release of Gaia, \textit{Gaia}-DR3 \citep{GaiaCollaboration2023}, while infrared photometry (JHKs) is taken from the Two Micron All Sky Survey \citep[2MASS][]{Skrutskie2006}. \texttt{StarHorse} then provides a posterior probability distribution for each of the output quantities. We consider the 50$^{th}$ percentile of this distribution as the parameters' best-fit value, with an associated Gaussian error equal to half of the 84$^{th}$ -- 16$^{th}$ percentiles interval.

The age sample is composed of only MSTO and SGB stars, based on their position in the Kiel diagram, following the selection presented in \citet{Queiroz2023}. These evolutionary stages represent the optimal ``sweet spot'' for age determination through isochrone fitting, as the shape of the curves varies significantly with age. Leveraging this variability, along with the high-quality stellar parameters from \citet{Guiglion2024}, and Gaia's very precise parallaxes, the sample achieves a mean statistical uncertainty of just 12\% in age and 1\% in distance.

In \citetalias{Nepal2024}, the set of isochrones used spans metallicities from [Fe/H] $=-$2.2 to $+$0.6 and ages from 0.025 to 13.73 Gyr, with the upper age limit set by the value of $t_U$ in a flat $\Lambda$CDM cosmology. This work is based on the same dataset and uses parameters derived with the same methodology. However, aiming to obtain estimates suitable for use in a cosmological context, independent of any prior assumptions, we recomputed the ages with \texttt{StarHorse}, extending the explored range from 0.025 to 20 Gyr.

\section{Analysis}
\label{sec:Analysis}
This section describes the selection process of an optimal age sample, the analysis of the systematic effects involved, and details the approach taken to identify the oldest stars in the dataset.

\subsection{The selection process}
\label{sec:selectionprocess}

The primary goal of this analysis is to obtain reliable, cosmology-independent age estimates. To achieve this, we implemented a rigorous selection process, as described below.

{\sffamily \textit{(1) Parent sample}}. From the full age sample of \citetalias{Nepal2024}, we selected stars older than 12.5 Gyr with age uncertainties below 1 Gyr (hereafter the parent sample), yielding 2,911 objects, about 10\% of the original sample. The 1 Gyr uncertainty threshold ensured high-quality measurements while still retaining a statistically robust dataset\footnote{We note that adopting a more stringent uncertainty cut of 0.75 Gyr would have reduced the parent sample to roughly one third of its current size, while decreasing the final age estimate by only 0.1 Gyr.}.
We underscore that this choice did not introduce any bias towards younger ages: even though a positive correlation between age and age error does exist, this holds only up to approximately 10 Gyr, after which errors remain comparable in the range 10 -- 15 Gyr. Conversely, in adopting this cut in precision, we found an overdensity at very old ages, due to solutions converging towards the edge of the prior, thus exhibiting artificially small errors \citepalias[see discussion in][]{Nepal2024}. This can be observed in the left panel of Fig. \ref{fig:3histo_kiel}, where the age distribution for the parent sample is shown in grey, revealing a clear double-peak in age: one around 13 Gyr and the other near 19 Gyr.

{\sffamily \textit{(2) Conservative MSTO-SGB stars cut}}. As anticipated in Sect.~\ref{sec:2DATA}, a cut in the Kiel diagram was already made in the age sample to select only MSTO and SGB stars. This selection, though, could be contaminated by stars in different evolutionary stages (low-luminosity giants, MS stars), hidden binaries, or mass-stripped stars (see Appendix \ref{appendix_conscut} for details). To this purpose, we adopted a more restrictive selection, already tested in \citetalias{Nepal2024}, in the log(g)--$\mathrm{T_{eff}}$ plane. This is shown with a solid line in the central panel in Fig. \ref{fig:3histo_kiel}, while the resulting sample, of 2148 objects, is shown in orange. This cut was highly effective in suppressing the 19 Gyr peak, suggesting that this overdensity was primarily due to contamination.

{\sffamily \textit{(3) Consistency of input--output parameters.}} Age estimates can also be affected by parameter degeneracies, which may introduce systematic biases. In particular, the age–metallicity degeneracy displayed a clear trend: more metal-poor stars ([M/H]$<-0.5$) were systematically shifted toward higher metallicities by 0.1–0.2 dex, while yielding old ages exceeding 18 Gyr. This is shown and discussed in more detail in Appendix \ref{appendix_samplesel}. To mitigate this effect, we imposed a constraint on the maximum allowed discrepancy between the input and output values of [M/H], the first being the results from \citet{Guiglion2024}, as anticipated in Sect. \ref{sec:1Intro}. We chose a symmetric cut, discarding the 5\% tails of the $\Delta$[M/H] distribution, namely $|\Delta$[M/H]$|<$0.025. Following the same logic, we imposed similar cuts on $\mathit{T_{eff}}$, $|\Delta T_{\textit{eff}}|<$30 K, $\mathit{log(g)}$, $|\Delta$log(g)$|<$0.05, and dust extinction, $|\Delta A_{V}|<$0.1 mag. This further reduced the sample to 1078 stars (see Fig. \ref{fig:3MH_age}).

{\sffamily \textit{(4) Removing strongly degenerate solutions.}} Next, we considered the symmetry of the posterior distributions in age and mass as quality indicators, as these two parameters are not constrained by Gaussian priors and are therefore more prone to asymmetries or anomalies in their posterior shapes. First, we discarded age measurements for which the asymmetry (difference between the upper and lower uncertainties) exceeded 0.1 Gyr. We then applied a Kolmogorov-Smirnov test to assess how well each posterior probability distribution function (PDF) conformed to a Gaussian, excluding all cases where the probability of deviation from Gaussianity exceeded 99.5\%. This reduced the sample to 297 stars. 

{\sffamily \textit{(5) Visual inspection.}} 
As a final step, we visually inspected the corner plots to exclude anomalous cases missed by previous cuts. This ``blind'' inspection, performed without viewing parameter values, focused on identifying asymmetries or double peaks in the posterior distributions. Based on PDF shape, stars were classified into three quality tiers: {\em great} (78 stars, symmetric and clean), {\em good} (107 stars, minor tail features), and {\em bad} (112 stars, significant asymmetries or almost double peaks), with the latter excluded from the final sample. Examples of this classification process are shown in Appendix \ref{appendix_visualinsp}.

\bigskip
Finally, we arrived at a refined final sample of 185 stars with precise age determinations, robust stellar parameters, and well-behaved posterior distributions. Within this set, we also defined a <<golden sample>> consisting of the 78 stars that exhibit the highest-quality PDFs. Both samples show a 5\% (stat) precision in age and 2\% (stat) precision in mass, on average.\\
The selection favours the more metal-rich part of the parent sample ([M/H] $> -0.5$, see right panel in Fig. \ref{fig:3histo_kiel}) with low $\alpha$-enhancement ([$\alpha$/Fe] $< 0.15$), mainly because it is where the \texttt{hybrid-CNN} delivers its most precise metallicity estimates -- key for reliable ages. Even though one would expect the oldest stars to be more metal-poor ([M/H]$<-1$), recent studies \citep[e.g.][and references therein]{trevisan_analysis_2011,anders_dissecting_2018,Miglio2021,Nepal2024,Borbolato2025} have already shown the existence of old, more metal-rich, low-alpha stars, compatible with a high-star formation rate and rapid metal enrichment in the early Milky Way. At high-redshift, while direct stellar observations are currently not possible, recent studies of gas phase metallicities have revealed a large dispersion, including super-solar estimates \citep[e.g.][]{Huyan2025,deepak_global_2025}. Most stars also show low extinction ($A_V < 0.5$ mag) as the whole final sample is confined within $\sim$700 pc from the Sun. 

\begin{figure}[h!]
    \centering
    \includegraphics[width=0.99\linewidth]{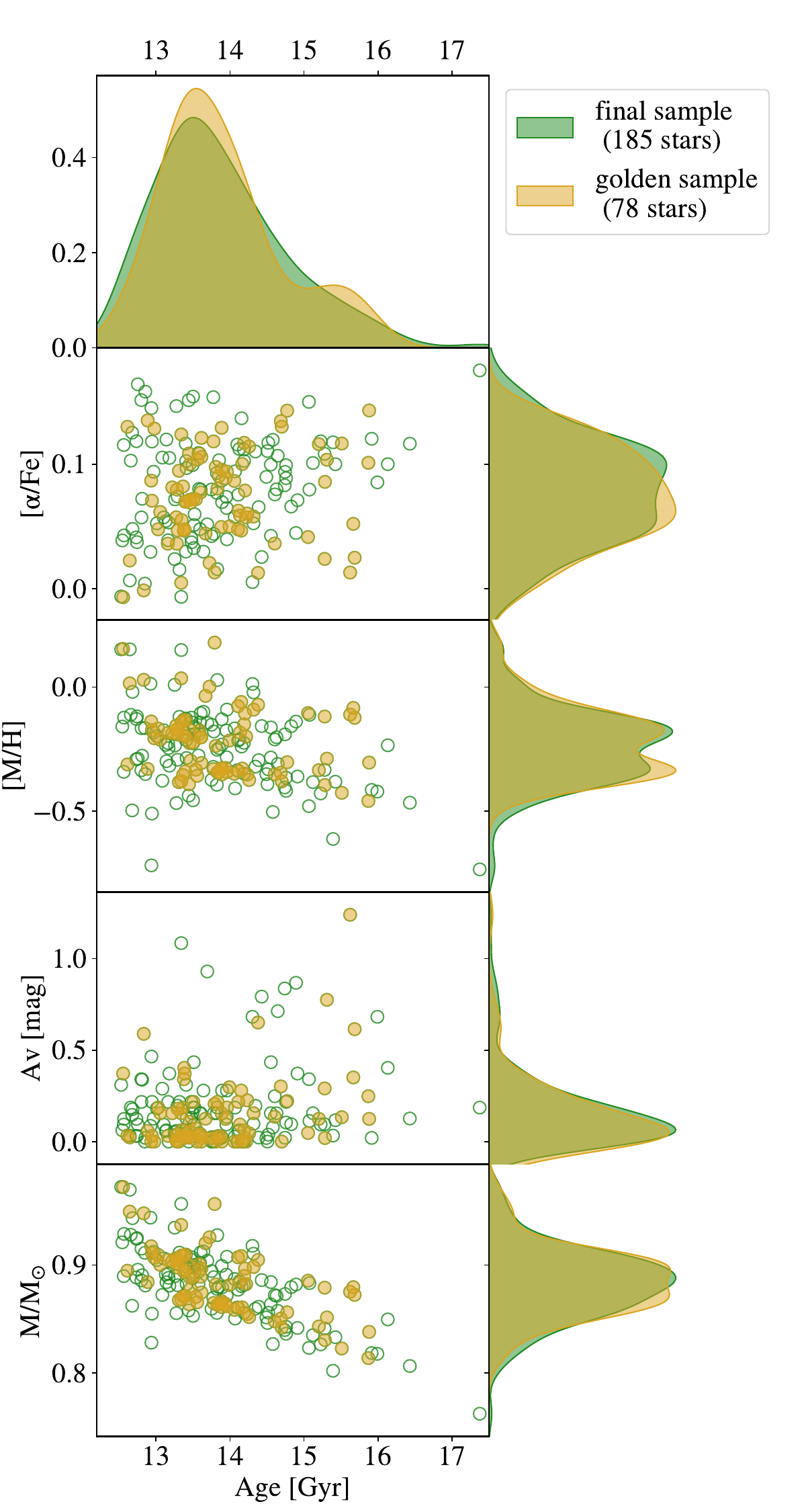}
    \caption{Trends with age of the main parameters, from the bottom: mass, dust reddening, overall metallicity and $\alpha$-enhancement. On the right, the corresponding normalised distributions are shown for each parameter, and at the top is the distribution in age.}
    \label{fig:3_final_golden}
\end{figure}
In Fig. \ref{fig:3_final_golden}, we show the distribution of the main physical parameters ($\alpha$-enhancement, metallicity, dust reddening, and mass) versus age for the final and golden sample. The histograms show that our sample does not deviate much from solar values. Considering the final sample, the mean and standard deviation for each quantity are:
\begin{align*}
\text{$\mathrm{\langle M/M_\odot\rangle}$} &= 0.88 \pm 0.03\\
\text{$\mathrm{\langle[\alpha/Fe]\rangle}$} &= 0.17 \pm 0.21\:\\ 
\text{$\mathrm{\langle A_V\rangle}$} &= 0.08 \pm 0.04 \text{\:mag}\\ 
\text{$\mathrm{\langle[M/H]\rangle}$} &= -0.24 \pm 0.15\\ 
\end{align*}

We also observe an inverse correlation between age and mass, with the least massive stars exhibiting the oldest ages. While this trend is expected, it may also hint at potential contamination, particularly in the low-mass tail, where a noticeable peak appears in the age distribution of the golden sample. These stars, with masses as low as 0.8 $M_\odot$, could represent remnants of originally more massive stars, evolved in binary systems that stripped away part of their gas. Recent evidence suggests that the fraction of mass transfer, degenerates, and mergers could represent about $\sim$20\% of Population II stars \citep{Fuhrmann2021}, so we will need to account for this component in the following.

\subsection{Systematic effects}\label{sec:systematics}

An important aspect to account for when relying on absolute age determination are the systematic effects at play. When performing isochrone fitting, two main sources of systematic uncertainties are involved: the first arising from the choice of the stellar models \citep[see, e.g.,][and references therein]{lebreton_how_2014,nsamba_asteroseismic_2021}, the second depending on potential biases affecting parameters like metallicity or $\alpha$-enhancement assumed in the fitting process.

In \citetalias{Nepal2024}, it is observed how the $\alpha$-elements abundance could be under-predicted for some MSTO stars by up to 0.08 dex, owing to a comparison with common GALAH DR3 \citep{Buder2021} stars. This little shift, if present, would cause an underestimation of the overall metallicity on which a Gaussian prior is assumed, thus overestimating the derived age. To account for this possible bias, following the approach adopted in \citetalias{Nepal2024}, we perturbed the assumed [$\alpha$/Fe] by $\pm$0.1, finding, on average, a shift of $\pm$0.28 Gyr. However, as we discussed before, most of our selected stars have rather low $\alpha$-enhancements (see Fig. \ref{fig:3_final_golden}).

In terms of stellar models, a positive note is that their difference is minimal for solar-like objects, as all models are calibrated to reproduce observations of the Sun, and our final sample shows properties that deviate very little from solar-like values. Nevertheless, different assumptions (e.g., on the initial helium fraction, mixing length, or treatment of diffusion) can lead to different results. \citet{Joyce2023}, in particular, identifies variations in the mixing length parameter ($\alpha_{\mathrm{ML}}$) as one of the dominant sources of systematic uncertainty in stellar age estimates, producing shifts of 1--2 Gyr for a sample of MSTO stars. Given our narrower parameter space and higher precision, we adopt the lower bound (1 Gyr) as a reasonable estimate of this systematic. Further details are discussed in Appendix \ref{appendix_systmodels}.

\subsection{Identifying the tail of spurious ages}
\label{sec:identifyingthetail}

As noted in Sect.~\ref{sec:Analysis}, the tail towards oldest ages could be contaminated by stars that lost some mass, causing them to appear artificially older, but also by undetected binaries with unequal mass, as shown in \citet{Woody2025}. Being close to the physical limit imposed by $t_U$, these spurious solutions should pop out as a secondary distribution at older ages. To test this, we fit a Gaussian Mixture Model to both the good and golden samples, allowing the data to determine the optimal number of components. Based on both the Bayesian and Akaike Information Criteria, the fit favoured two populations: a narrow peak at 13.4$\pm$0.8 Gyr and a broader one at 14.8$\pm$1.5 Gyr. Given the presence of this secondary component, we estimated the contamination fraction with a hierarchical Bayesian model implemented through \texttt{PyMC} (details are provided in Appendix \ref{appendix_tail}).
The fit found a tail of about 11\% contaminants, peaking near 14.3 Gyr with a dispersion of 0.8 Gyr. Conservatively, we excluded stars with >20\% contamination probability, 25 in total (11 in the golden sample). The cumulative PDFs of the clean final and golden sample peak at 13.6$\pm$1.0 Gyr and 13.6$\pm$0.9 Gyr, respectively, confirming no significant difference between the two samples. Hence, we adopted the clean final sample (160 stars) as our reference.

\section{Implications for cosmology}\label{sec:4CosmoAnalysis}

\begin{figure*}[h!]
  \centering
  \begin{minipage}{0.3\textwidth}
    \caption{Cumulative posterior distribution in age for final and golden sample. The distributions including the systematic component of the error are shown with solid lines in the same colours. The upper axis shows the corresponding $H_0$ value, assuming $\mathit{z_f}$=20. In the lower panel, the age ranges covered by the stars in the final and golden samples and their means are shown in comparison with the oldest ($>12.5$ Gyr) GCs in \citet{Valcin2025} and the oldest bulge GCs in \citet{Souza2024}.}
    \label{fig:age_h0_dist}
  \end{minipage}
  \hspace{0.5cm}
  \begin{minipage}{0.65\textwidth}
    \includegraphics[width=\linewidth]{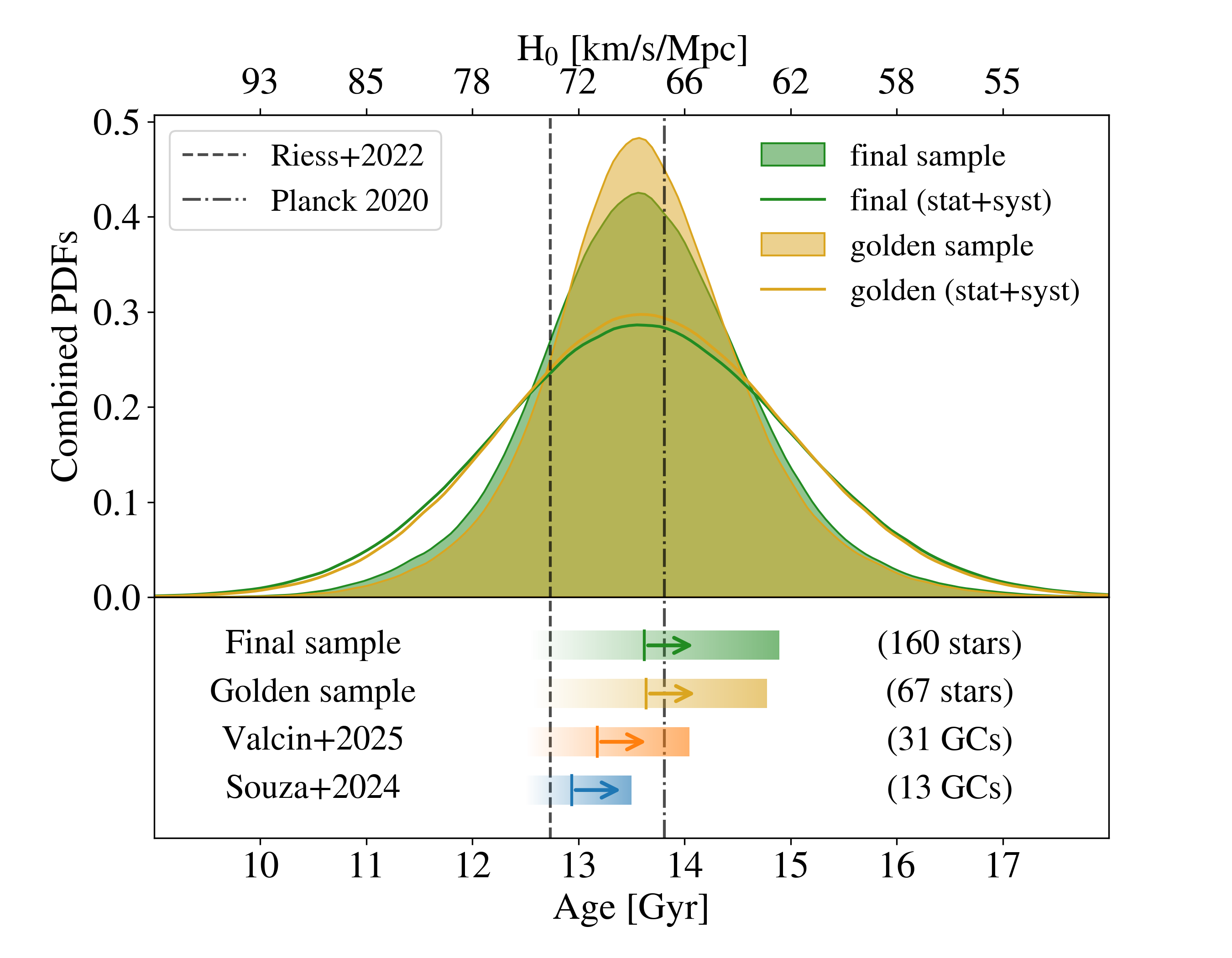}
  \end{minipage}%
\end{figure*}


\subsection{From the stellar ages to the age of the Universe}

To convert the stellar ages into constraints on $t_U$ we need to take into account the delay, $\delta t$, between the Big Bang and the moment these objects formed. As the ages of the oldest objects represent a lower limit to $t_U$, a conservative choice would imply accounting for the smallest possible value for $\delta t$. Theoretical models of stellar formation \citep{Galli2013,Bromm2011} and spectroscopically confirmed observations of the most distant galaxies \citep{CurtisLake2023,Carniani2024} show that the very first stars have formed at z$\gtrsim$11--14, and not earlier than z$\sim$20--30, the expected redshift of formation for Pop III stars. This corresponds to an interval $\delta t$ of about 0.2--0.4 Gyr.

In Fig. \ref{fig:age_h0_dist}, we show the cumulative PDF of our final and golden sample. With lines in the same colours, we report the cumulative distribution obtained by adding the systematic errors in quadrature to each star, as discussed in Sect. \ref{sec:systematics}. The final cumulative PDF, shows a mean and standard deviation of: 
\begin{align*}
\text{age} &= 13.6 \pm 1.0\:\text{(stat)} \pm 1.3\:\text{(syst) } \text{Gyr}.
\end{align*}
Accounting also for the minimum formation time for these stars, this produces a lower limit to the age of the Universe:
\begin{align*}
    \mathrm{t_U \geq (13.8 - 14.0)} \pm 1.0\:\text{(stat)} \pm 1.3\:\text{(syst) } \text{Gyr},
\end{align*}
which would increase, should these stars have formed at z$<$11.

In the upper axis of Fig. \ref{fig:age_h0_dist}, the values of the Hubble constant $H_0$ are also reported assuming a redshift of formation for the sample $z_f=20$. Under these assumptions, the lower limit on the age of the Universe translates, in terms of $H_0$, into:
\begin{align*}
    \mathrm{H_0 \leq 68.3^{+5.4}_{-4.7}\: (stat)\: ^{+7.2}_{-5.9}\: (syst)\: \text{km/s/Mpc}},
\end{align*}
and would further decrease by 1.2 km/s/Mpc assuming $z_f=11$.
For comparison, the two values currently leading the Hubble tension, \citet{PlanckCollaboration2020} and \citet{Riess2022}, are also shown with dashed lines. We underline that the assumptions made here to connect $H_0$ and $t_U$ are commonly adopted, but represent a particular case. The relation between $H_0$ and $t_U$ can be influenced by different factors, especially the value assumed for $\Omega_m$, but also by the cosmological model in general. In Sect. \ref{alternative_cosmo}, we discuss this aspect in more detail.
Regardless of the implications on the value of $H_0$, the sample identified in this work provides important and direct constraints on the age of the Universe itself, representing an observational anchor point to any cosmological model.

Considering the single PDFs of the 160 stars in the final sample (excluding, for now, systematic uncertainties) and shifting them by the minimum possible delay (0.2 Gyr), we find that at the 90\% confidence level (CL), 70 stars indicate an age of the Universe older than 13 Gyr, and 29 stars suggest an age older than 13.5 Gyr, while no star exceeds 14.1 Gyr at 90\% CL. Notably, in order for the data to indicate a significant drop in ages at 13 Gyr or younger, the full systematic error budget would be needed, consistently shifting the results towards younger ages. While this may be plausible for the [$\alpha$/Fe] estimation component (0.3 Gyr in the error budget), there is no evidence favouring stellar models that predict systematically younger ages. On the contrary, considering models with higher $\alpha_{ML}$ (see Appendix \ref{appendix_systmodels} for details), age estimates could increase by up to 1 Gyr.

The results of this work are also complemented and supported by independent age estimations obtained for very old GCs, as the ones in \citet{Valcin2025} and the bulge GCs in \citet{Souza2024}. In the lower panel of Fig.~\ref{fig:age_h0_dist}, we show the age ranges and mean values from both studies restricted to clusters older than 12.5 Gyr, consistent with our selection. The figure highlights how the tail of the oldest GCs overlaps with the age distribution of our sample. Although the stellar models adopted in those studies differ from those used here, the age ranges of the oldest GCs vary by less than 1 Gyr, with average ages differing by only $\sim$0.5\footnote{Peculiar GCs showing a spread in metallicity (namely NGC5286, NGC5139, and NGC7089) were removed from the sample.} Gyr \citep{Valcin2025} and $\sim$0.7 Gyr \citep{Souza2024}. Remarkably, both samples show metallicities comparable to those in our sample, reinforcing the conclusion that such old ages are achievable even at these metallicities, supporting a scenario of rapid early formation.

\begin{figure*}
    \centering
    \includegraphics[width=.95\linewidth]{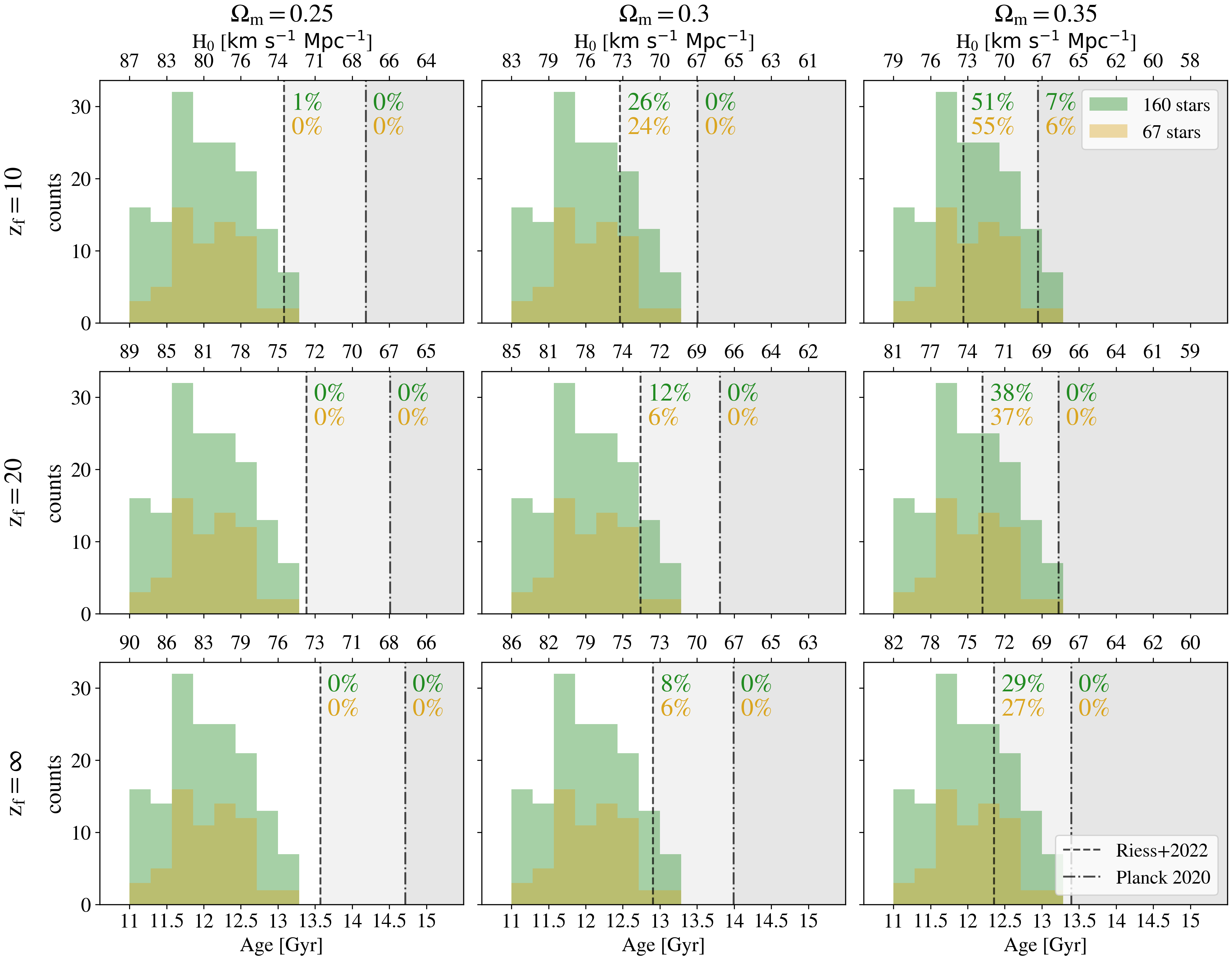}
    \caption{Distribution of the $10^{th}$ percentile in age for the clean final (green) and golden (gold) samples. In each panel, the top axis shows the corresponding $H_0$ values assuming a flat $\Lambda$CDM and a different value of $\Omega_m$ and $z_f$: rows from top to bottom set $z_f=10, 20, \infty$, columns from left to right fix $\Omega_m =$ 0.25, 0.3, 0.35. The dashed and dashed-dotted lines report the $H_0$ measurements from \citet{Riess2022} and \citet{PlanckCollaboration2020}, respectively. Next to each one of them, the percentage of stars in the final (green) and golden (gold) samples pointing to a lower $H_0$ at 90\% CL (stat+syst) is reported.} 
    \label{fig:age_h0_grid}
\end{figure*}

\subsection{Alternative cosmological assumptions}\label{alternative_cosmo}

In Fig. \ref{fig:age_h0_grid} we show how the upper axis of Fig. \ref{fig:age_h0_dist} changes when changing the values of the matter density parameter $\Omega_{M}$ and of the assumed redshift of formation for these stars, $z_f$. The histograms show the $10^{th}$ percentile distribution in age of the final and golden samples, including the systematic component of the error for each star. All panels share the same lower horizontal age axis, while the upper one, in $H_0$, varies depending on the assumed value of $\Omega_m$ and $z_F$. This allows us to show how the distribution found in this work compares with the measurements from \citet{PlanckCollaboration2020} and \citet{Riess2022}, represented by a dashed and a dashed-dotted line, respectively, at varying cosmological assumptions. The percentages reported in each panel, next to these lines, show the fraction of stars older than the respective ages of the Universe at 90\% CL (stat+syst), thus pointing to a lower $H_0$ value at 90\% CL. 

The setting adopted in Fig. \ref{fig:age_h0_dist} corresponds to the one in the central panel in Fig. \ref{fig:age_h0_grid}. Overall, this shows how $z_f$ and $\Omega_m$ act when translating the age of the oldest objects into $H_0$: the higher $z_f$ is assumed, the higher $H_0$ is retrieved, while the opposite is true for $\Omega_m$. In the context of the Hubble tension, this shows that the sample in this work points to $H_0$ lower than the CMB value only when $\Omega_m = 0.35$ and $z_f=10$. Compared to \citet{Riess2022}, instead, all configurations with $\Omega_m\geq0.3$ show at least 6\% of stars, up to 51\%, pointing to a lower value of $H_0$ at 90\% CL. The only configurations showing no tensions with any of the two measurements are the ones with $\Omega_m = 0.25$.

\section{Conclusions}\label{sec:5CONCLUSIONS}

In this paper, the ages of the oldest stars from Gaia DR3 are used to constrain the age of the Universe, $t_U$, representing the first attempt to use the ages of single stars as cosmic clocks with a statistically significant sample.

We considered the $\sim$200,000 stars from \citetalias{Nepal2024}, with ages and masses estimated via the Bayesian code \texttt{StarHorse}, but allowing ages to vary up to 20 Gyr without a cosmological prior. We selected the $\sim$3,000 stars older than 12.5 Gyr with age uncertainties below 1 Gyr. Through a careful selection process including cuts in the Kiel diagram, stellar parameter quality, symmetry of posterior distributions, and a final visual inspection, we removed stars with potentially biased age estimates, especially those skewed to older values, to obtain a conservative and robust lower limit on stellar ages and thus on $t_U$. 

We identified two sub-populations in age, one at $\sim$13.7 with very little dispersion ($\sim$0.3 Gyr) and an older one, at $\sim$14.8 Gyr, with a dispersion of $\sim$0.8 Gyr. Assuming a fraction of stars could be composed of older contaminants (e.g., mass-stripped stars, binaries), appearing older than they are, we conservatively excluded all stars belonging to the second peak ($\sim11\%)$.

The final sample counts 160 stars, with a cumulative posterior distribution peaking at 13.6 $\pm$1.0 (stat) $\pm$1.3 (syst) Gyr. The main source of systematic error comes from stellar models. Accounting for the minimum possible delay between the Big Bang and the formation of these stars, 0.2 Gyr at $\mathit{z_f}=20$, we derive a conservative lower limit on $t_U$, and an upper limit on $H_0$:
\begin{align*}
\text{age} &\geq 13.8 \pm 1.0\,\text{(stat)} \pm 1.3\,\text{(syst)}\ \text{Gyr}, \\
H_0 &\leq 68.3^{+5.4}_{-4.7}\,\text{(stat)}\,^{+7.2}_{-5.9}\,\text{(syst)}\ \text{km\,s}^{-1}\,\text{Mpc}^{-1}.
\end{align*}

Considering them one by one, 70 stars indicate an age of the Universe older than 13 Gyr, while no star exceeds 14.1 Gyr at 90\% CL (stat). Notably, the full systematic error budget would be needed, consistently pointing towards younger ages, to move this drop to 13 Gyr or less.

In conclusion, this work shows how the ages of single stars derived from isochrone fitting can provide stringent constraints on $t_U$, and a robust anchor point to any cosmological model. While this represents a significant first step, future data releases from \textit{Gaia} will enable similar analyses on larger stellar samples with improved precision. Furthermore, the accuracy and reliability of age determinations can be improved by obtaining metallicities from high resolution spectroscopy, minimising, at the same time, the systematic error due to the $\alpha$-enrichment. However, only with missions such as Haydn \citep{miglio_haydn_2021} will it be possible to achieve accurate ages for field stars in the MW.

\begin{acknowledgements}
    We thank Andrea Miglio, Arman Khalatyan, and the e-science department of AIP for their contributions to this paper. Part of this project was conducted at AIP within the Marco Polo program.
    ET acknowledges COST Action CA21136 – “Addressing observational tensions in cosmology with systematics and fundamental physics (CosmoVerse)”, supported by COST (European Cooperation in Science and Technology). CC acknowledges the Astronomy Department of the University of S\~{a}o Paulo and FAPESP. MM acknowledges the financial contribution from the grant PRIN-MUR 2022 2022NY2ZRS 001 “Optimizing the extraction of cosmological information from Large Scale Structure analysis in view of the next large spectroscopic surveys” supported by Next Generation EU. MM and AC acknowledge support from the grant ASI n. 2024-10-HH.0 “Attività scientifiche per la missione Euclid – fase E”. This work was partially funded by the Spanish MICIN/AEI/10.13039/501100011033 and by the ``ERDF A way of making Europe'' funds by the European Union through grant RTI2018-095076-B-C21 and PID2021-122842OB-C21, and the Institute of Cosmos Sciences University of Barcelona (ICCUB, Unidad de Excelencia ’Mar\'{\i}a de Maeztu’) through grant CEX2019-000918-M. FA acknowledges financial support from MCIN/AEI/10.13039/501100011033 through a RYC2021-031638-I grant co-funded by the European Union NextGenerationEU/PRTR.
    
\end{acknowledgements}

\bibliographystyle{aa}
\bibliography{references}
\newpage
\begin{appendix}

\section{Sample selection}\label{appendix_samplesel}

\subsection{Conservative cut in the Kiel diagram}\label{appendix_conscut}

Following what was already presented in \citetalias{Nepal2024}, we adopted the following cut in $\mathit{log(g)-T_{eff}}$ plane:
\begin{equation}
    \begin{cases}
        \mathrm{log(g) < 4.1}\\
        \mathrm{log(g) > -0.0003 \times T_{eff}+4.8}\\
        \mathrm{T_{eff} > 500 \times log(g)+3000}
    \end{cases},
\end{equation}
selecting a restricted area in the Kiel diagram. The cut in $\mathit{log(g)}<4.1$, in particular, is what impacted most on the parent sample and removed most of the stars older than 18 Gyr. Above this threshold, as is clear from the central panel in Fig. \ref{fig:3histo_kiel}, we would select stars that overlap with the turn-off, but potentially still in MS. The simple scatter of MS stars, but also unresolved binary systems\footnote{Wide binaries are already excluded in the original selection from \citetalias{Nepal2024}.}, could imply a lower surface gravity \citep[see, e.g., ][for a discussion on binarity]{PriceWhelan2020,Anders2022}. This has a different effect on the resulting age depending on the position in the diagram: above $log(g)\sim4.1$, around the turn-off, this can drag the age to much older values because isochrones of different age around the MS are very close by; below $log(g)\sim4.1$, above the turn-off, instead, such upwards shift would typically result in a younger age, but with a much less dramatic effect because the isochrones of different age are distinct parallel lines after the turn-off. We underline that by selecting this region of the Kiel diagram, we are not removing by definition the oldest solutions: isochrones of ages up to 20 Gyr are still included in the selected area, as visible in the central panel of Fig. \ref{fig:3histo_kiel}.

\subsection{Consistency of input--output metallicity}

\begin{figure}[h!]
    \centering
    \includegraphics[width=\linewidth]{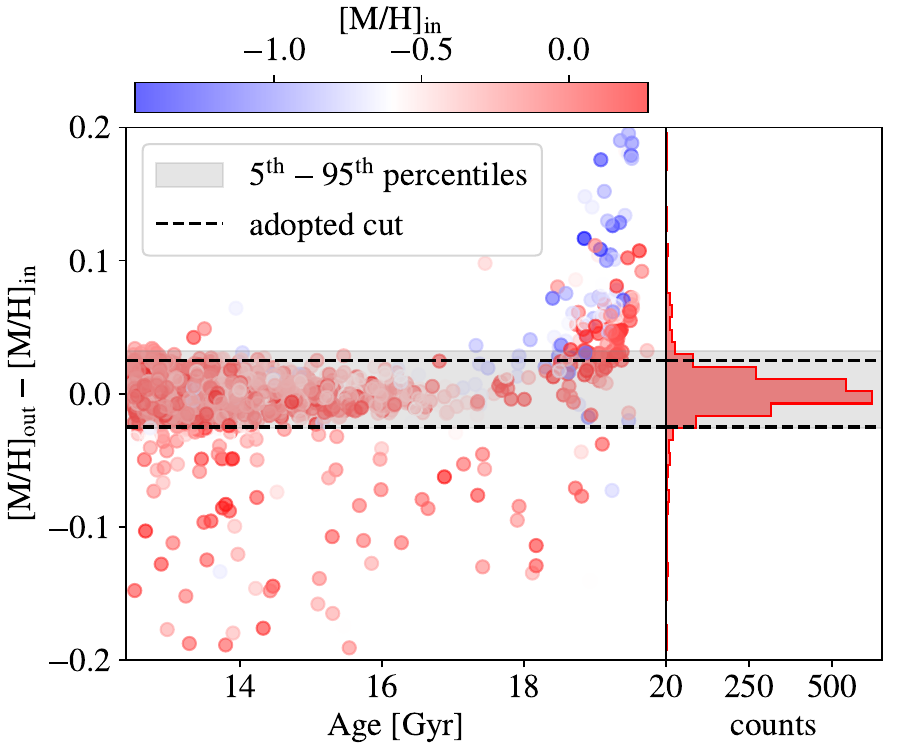}
    \caption{Discrepancy in [M/H] between StarHorse's output and the measurements from \citet{Guiglion2024}, used as a prior.}
    \label{fig:3MH_age}
\end{figure}

In Figure \ref{fig:3MH_age}, the difference is shown between the \texttt{StarHorse} overall metallicities and the values from \citet{Guiglion2024}, used as priors, for the sample selected after the conservative cut (2148 stars).
It clearly highlights how the oldest ages, above 18 Gyr, are all related to an overestimation of the input metallicity by up to 0.2 dex. From the histogram on the right, it is also clear that these stars are just a very small fraction, less than 5\% of the entire population. The dashed lines show the cut that we adopted to remove these stars and keep only the most stable results, in which the code did not deviate much from the metallicity prior.

%

\subsection{Visual inspection}\label{appendix_visualinsp}
\begin{figure}[h!]
    \centering
    \subfigure[Example of the best PDF]{
        \includegraphics[width=0.35\textwidth]{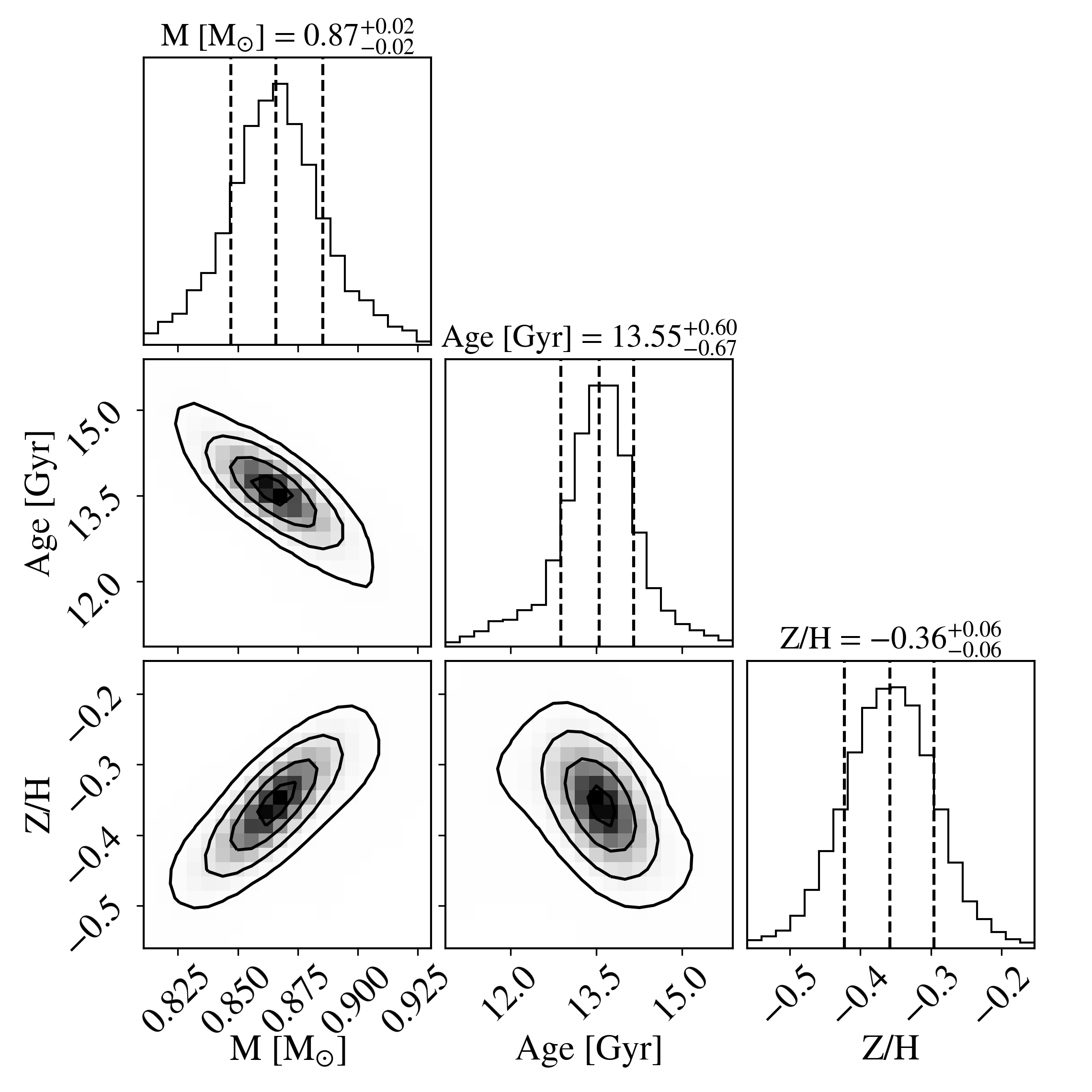}
    }
    \hspace{0.01\textwidth}
    \subfigure[Example of good PDF]{
        \includegraphics[width=0.35\textwidth]{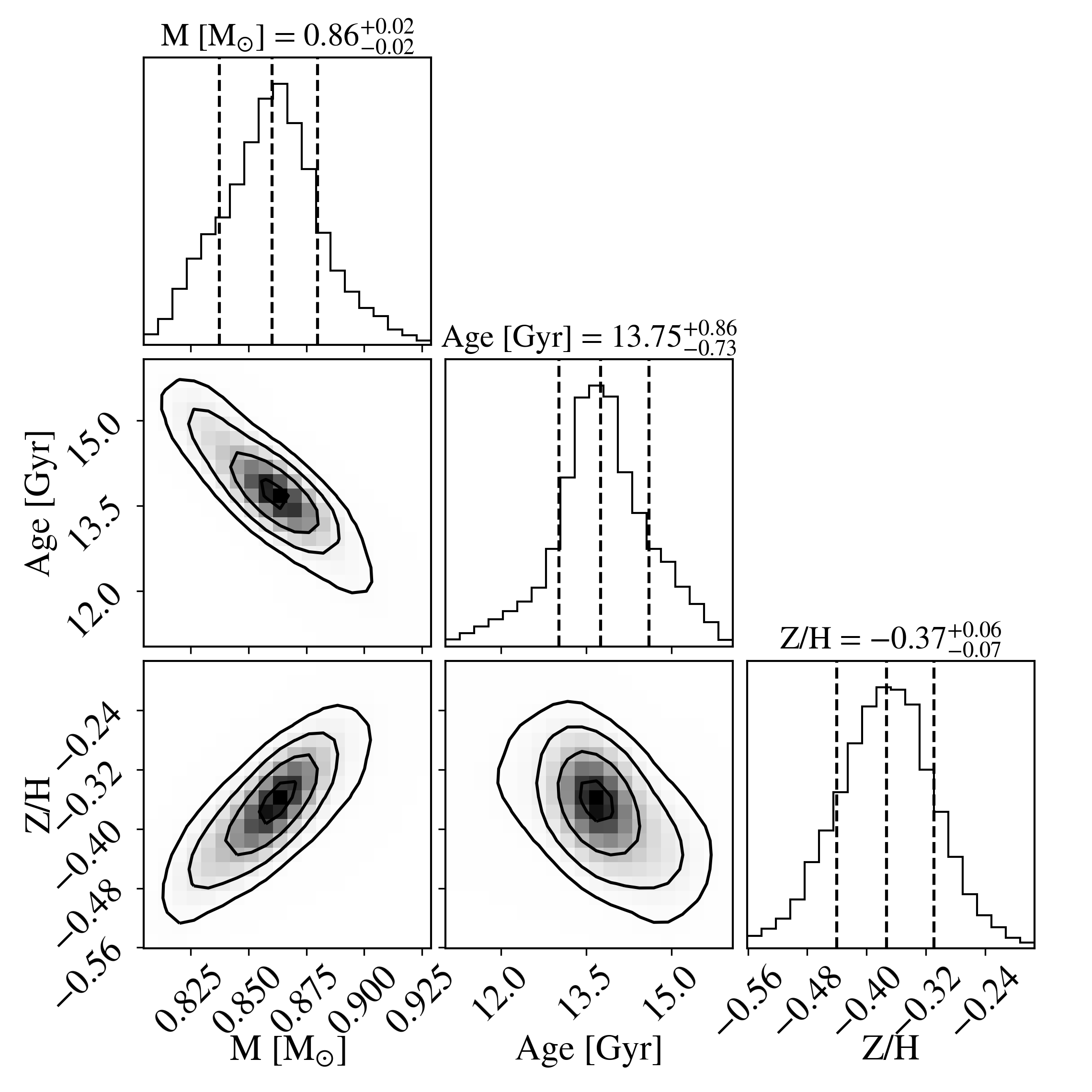}
    }
    \hspace{0.01\textwidth}
    \subfigure[Example of bad PDF]{
        \includegraphics[width=0.35\textwidth]{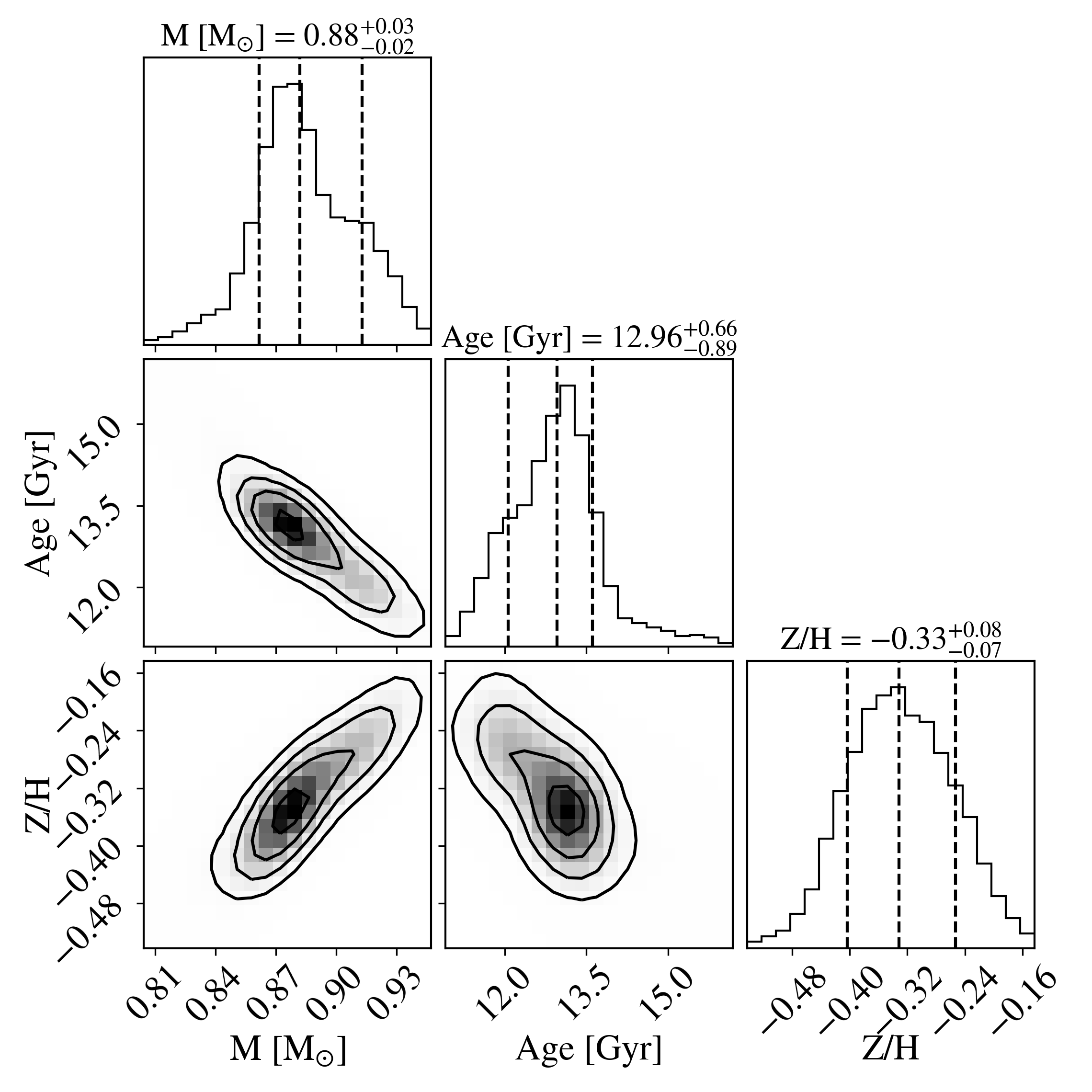}
    }
    \caption{Examples of corner plots of different quality, classified in the visual inspection phase into best, good, and bad PDFs.}
    \label{fig:best_good_bad}
\end{figure}
In Fig.~\ref{fig:best_good_bad}, three corner plots are shown for stars representing different levels of quality in their posterior probability distribution functions (PDFs) for mass and age. 
The top panel illustrates one of the best cases, characterized by a Gaussian and symmetric distribution in both parameters. The middle panel presents a good-quality PDF, with a single peak and an overall Gaussian-like shape, but slight asymmetries appear in the tails beyond the 1$\sigma$ range. The bottom panel shows a poor-quality fit that was excluded from the final sample. Although the main peak is visible, a secondary peak appears at higher masses and lower ages, close enough to the main one that it was not flagged in earlier selection steps. This bimodality suggests a non-negligible probability assigned to a different solution, reducing the robustness of the fit for the purposes of this study.


\section{Systematics: stellar models}\label{appendix_systmodels}

To derive a realistic estimation of the systematic component of the error introduced by the stellar models, we followed the approach presented in \citep{Joyce2023}. The authors focus on the impact of the convective mixing length parameter, $\mathit{\alpha_{ML}}$, that they argue is likely dominating the systematic error budget linked to the stellar models, and they vary this parameter on a wide range -- 1.4 to 2.3 when considering MSTO and SGB \citep[see Fig. 11 in ][]{Joyce2023}. To mimic this variation, they perform a Monte Carlo simulation perturbing the set of isochrones by 200 K in $\mathit{T_{eff}}$ and 0.17 dex in log(g), on average, finding an average shift in ages of 1-2 Gyr, where a higher $\mathit{\alpha_{ML}}$ produces older ages. In order to adapt this analysis to our case, we need to make three considerations.\\
First, the range adopted for $\mathit{\alpha_{ML}}$ is very wide if compared to the typically used values in most stellar models, ranging mostly from 1.6 to 1.9 \citep[see, e.g., Table 4 in ][ for a summary]{Amard2019}. The models used in this work, PARSEC, set $\mathit{\alpha_{ML}}=1.74-1.77$, which is in the middle of this interval. \\
Second, the Kiel diagram area covered by the final sample in this work is confined between 5400-5700 K in $\mathit{T_{eff}}$ and 3.8-4.1 in log(g), where the effect of $\mathit{\mathit{\alpha_{ML}}}$ is minimal and fully horizontal (i.e., it produces only a shift in $\mathit{T_{eff}}$). Considering the full range 1.4-2.3, this produces shifts of $\pm$100-150 K from the mean value, while considering the most used range 1.6-1.9, this implies fluctuations of $\pm$30-50 K.\\
Third, the uncertainties in $\mathit{T_{eff}}$, log(g), and metallicity in this work are up to 4-5 times smaller than the ones considered in \citep{Joyce2023}.\\
Owing to these considerations, we conclude that the lower bound of the interval identified in \citet{Joyce2023} -- 1 Gyr -- is an overly conservative choice to account for the systematic component of the error related to the stellar models in this work.

Systematic uncertainties may also arise from discrepancies between the actual stellar helium content and the values assumed in model grids \citep[see][for a detailed discussion]{lebreton_how_2014,nsamba_asteroseismic_2021}. These studies show that adopting different assumptions for the initial helium content -- either reading it as a free parameter or rising it via a $\Delta$Y/$\Delta$Z relation -- can introduce systematic effects of the order of a few percent in mass and, therefore, in age. However, for our sample, most stars have metallicities close to solar, so their helium abundances are expected to be similar to those assumed in standard stellar models (solar mixture).

Another potential systematic effect linked to the stellar models is the one of atomic diffusion, causing the exterior of the star to appear more metal-poor with respect the internal, initial value of metallicity \citep[see, e.g.,][]{Korn2007}. The effect of diffusion is, anyway, already included in the PARSEC models and, even if the current knowledge of this process is still incomplete, it should be less and less relevant moving towards older ages and to metallicities around solar \citep[see Fig. 3 in ][]{Dotter2017}. For this reason, we decided not to add another source of systematic error owing to this effect.

\begin{figure*}
    \centering
    \includegraphics[width=.85\linewidth]{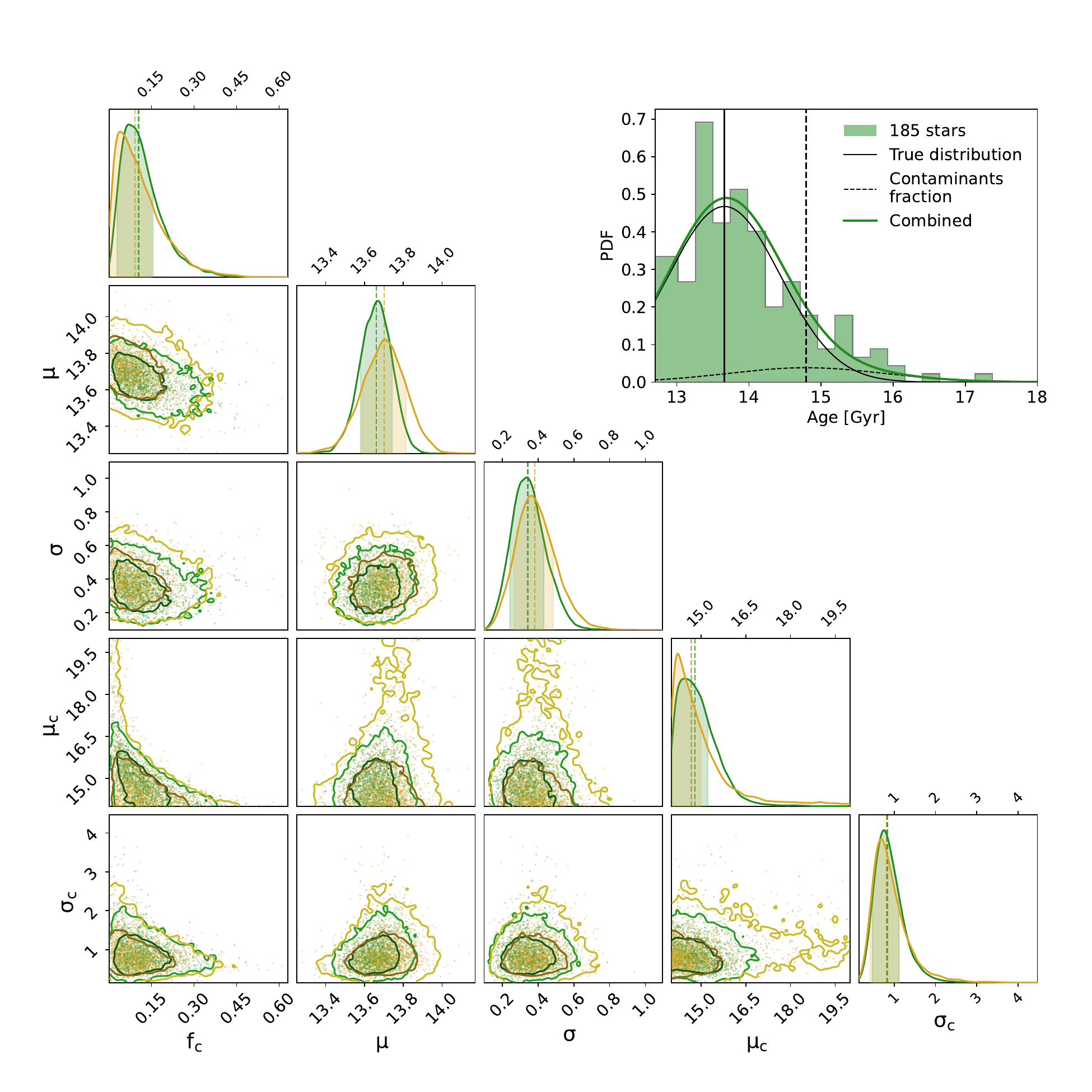}
    \caption{Joint posterior distribution of the five parameters included in the model reproducing the age distribution. Contours obtained with the final sample are shown in green, and in yellow for the golden sample. The inset shows the age distribution of the final sample and the two components resulting from the fit.}
    \label{fig:corner_plot}
\end{figure*}

\section{Removing potential contaminants}\label{appendix_tail}
In order to identify the possible contamination fraction of older stars that lost their mass, or undetected binaries, we fit our distribution as the combination of a true and a contaminant population, following the approach described in detail in \citet{Miglio2021}. 

Assuming a combination of two normally distributed samples, the first with mean $\mu$ and intrinsic dispersion $\sigma$, the second represented by a fraction, $f_c$, with mean $\mu_c$ and dispersion $\sigma_c$, the model can be summarised by the following likelihood function:
\begin{equation}
\begin{aligned}
\mathcal{L}(\mathbf{\theta} \mid \mathbf{x,\sigma_x})
= \prod_{i=1}^N
   &\Bigl[
     (1-f_c)\,\mathcal{TN}\bigl(\mu,\;\sigma^2 + \sigma_{x,i}^2,\,12,\,20\bigr)\\
&\quad
     +\,f_c\,\mathcal{TN}\bigl(\mu_c,\;\sigma_c^2 + \sigma_{x,i}^2,\,12,\,20\bigr)
   \Bigr],
\end{aligned}
\end{equation}
where $\mathcal{TN}(\mu, \sigma^2, a, b)$ is a normal distribution truncated in the range [a,b] and $(\mathbf{x, \sigma_x})$ is the set of age measurements and associated errors.\\
We adopt uniform, wide, priors on $\mu$ and $\mu_c$, respectively in the range 0--20 Gyr and 13--20 Gyr. On $\sigma$ and $\sigma_c$, we adopt lognormal priors, centred at 1 Gyr with standard deviation 0.5 Gyr. For the contamination fraction, $f_c$, we use a $\mathcal{\beta}$-function with parameters $a=2$ and $b=8$.\\
For the sampling of the posterior probability distribution, performed in \texttt{PyMC} with the No-U-Turn Samples (NUTS), we use four chains of 2000 steps, after performing 1000 burn-in steps, totalling 8000 samples.

The joint posterior distribution for the five parameters is shown in Fig. \ref{fig:corner_plot}, where the inset shows the decomposition in the final sample distribution. The median values and 68\% contours for each parameter are the following:
\[
\begin{array}{r@{\quad}l}
\text{Final sample} & \left\{
\begin{array}{l}
f_c = 0.10_{-0.05}^{+0.08} \\
\mu = 13.66_{-0.08}^{+0.08},\quad \sigma = 0.34_{-0.09}^{+0.10} \\
\mu_c = 14.79_{-0.52}^{+0.72},\quad \sigma_c = 0.83_{-0.28}^{+0.37}
\end{array}
\right. \\[1cm]
\text{Golden sample} & \left\{
\begin{array}{l}
f_c = 0.09_{-0.06}^{+0.10} \\
\mu = 13.70_{-0.11}^{+0.11},\quad \sigma = 0.38_{-0.10}^{+0.12} \\
\mu_c = 14.67_{-0.49}^{+1.03},\quad \sigma_c = 0.82_{-0.28}^{+0.47}
\end{array}
\right.
\end{array}
\]

We then removed from the sample all stars with a probability of being contaminants higher than 20\%, computed as follows:
\begin{equation}
    \mathrm{P(i=contam | x_i) = \frac{f_c\: \mathcal{N}(x_i;\mu_c,\sigma_c)}{f_c\: \mathcal{N}(x_i;\mu_c,\sigma_c) + (1-f_c) \mathcal{N}(x_i;\mu_t,\sigma_t)},}
\end{equation}
where $x_i$ is a single age measurement, and $\mathcal{N}(a;b,c)$ is the value of a normal distribution with mean $b$ and sigma $c$ evaluated at $a$. For each star, when possible, we computed this probability using the best-fit values obtained from fitting both the final sample and the golden sample, as listed in Fig. \ref{fig:corner_plot}. This process led us to discard 11 stars from the golden sample and 25 stars from the final sample, with all contaminants present in the first also identified in the second.

\end{appendix}

\end{document}